\documentclass[pra,reprint,floatfix]{revtex4-2}
\usepackage[breaklinks=true]{hyperref} 
\usepackage{graphicx}
\usepackage{siunitx}
\usepackage{physics}
\usepackage{bbm}
\newcommand{\A}{\mathrm{A}}
\newcommand{\B}{\mathrm{B}}
\newcommand{\Ss}{\mathrm{I}}
\newcommand{\AB}{\mathrm{AB}}

\newcommand{\HF}{\mathrm{HF}}
\DeclareSIUnit{\au}{{a.u.}}
\DeclareSIUnit\angstrom{\text {Å}}
\usepackage[capitalize]{cleveref}

\makeatletter
\let\save@mathaccent\mathaccent
\newcommand*\if@single[3]{%
  \setbox0\hbox{${\mathaccent"0362{#1}}^H$}%
  \setbox2\hbox{${\mathaccent"0362{\kern0pt#1}}^H$}%
  \ifdim\ht0=\ht2 #3\else #2\fi
  }
\newcommand*\rel@kern[1]{\kern#1\dimexpr\macc@kerna}
\newcommand*\widebar[1]{\@ifnextchar^{{\wide@bar{#1}{0}}}{\wide@bar{#1}{1}}}
\newcommand*\wide@bar[2]{\if@single{#1}{\wide@bar@{#1}{#2}{1}}{\wide@bar@{#1}{#2}{2}}}
\newcommand*\wide@bar@[3]{%
  \begingroup
  \def\mathaccent##1##2{%
    \let\mathaccent\save@mathaccent
    \if#32 \let\macc@nucleus\first@char \fi
    \setbox\z@\hbox{$\macc@style{\macc@nucleus}_{}$}%
    \setbox\tw@\hbox{$\macc@style{\macc@nucleus}{}_{}$}%
    \dimen@\wd\tw@
    \advance\dimen@-\wd\z@
    \divide\dimen@ 3
    \@tempdima\wd\tw@
    \advance\@tempdima-\scriptspace
    \divide\@tempdima 10
    \advance\dimen@-\@tempdima
    \ifdim\dimen@>\z@ \dimen@0pt\fi
    \rel@kern{0.6}\kern-\dimen@
    \if#31
      \overline{\rel@kern{-0.6}\kern\dimen@\macc@nucleus\rel@kern{0.4}\kern\dimen@}%
      \advance\dimen@0.4\dimexpr\macc@kerna
      \let\final@kern#2%
      \ifdim\dimen@<\z@ \let\final@kern1\fi
      \if\final@kern1 \kern-\dimen@\fi
    \else
      \overline{\rel@kern{-0.6}\kern\dimen@#1}%
    \fi
  }%
  \macc@depth\@ne
  \let\math@bgroup\@empty \let\math@egroup\macc@set@skewchar
  \mathsurround\z@ \frozen@everymath{\mathgroup\macc@group\relax}%
  \macc@set@skewchar\relax
  \let\mathaccentV\macc@nested@a
  \if#31
    \macc@nested@a\relax111{#1}%
  \else
    \def\gobble@till@marker##1\endmarker{}%
    \futurelet\first@char\gobble@till@marker#1\endmarker
    \ifcat\noexpand\first@char A\else
      \def\first@char{}%
    \fi
    \macc@nested@a\relax111{\first@char}%
  \fi
  \endgroup
}
\makeatother

\begin{document}

\title{Comparing real-time coupled cluster methods through simulation of collective Rabi oscillations}

\author{Andreas S. \surname{Skeidsvoll}}
\affiliation{%
Department of Chemistry, Norwegian University of Science and Technology, 7491 Trondheim, Norway
}%

\author{Henrik \surname{Koch}}
\email{henrik.koch@sns.it}
\affiliation{%
Scuola Normale Superiore, Piazza dei Cavalieri, 7, I-56126, Pisa, Italy
}%
\affiliation{ 
Department of Chemistry, Norwegian University of Science and Technology, 7491 Trondheim, Norway
}%

\date{\today}

\begin{abstract}
The time-dependent equation-of-motion coupled cluster (TD-EOM-CC) and time-dependent coupled cluster (TDCC) methods are compared by simulating Rabi oscillations for different numbers of non-interacting atoms in a classical electromagnetic field. While the TD-EOM-CC simulations are numerically stable, the oscillating time-dependent energy scales unreasonably with the number of subsystems resonant with the field. The TDCC simulations give the correct scaling of the time-dependent energy in the initial stages of the Rabi cycle, but the numerical solution breaks down when the multi-atom system approaches complete population inversion. We present a general theoretical framework in which the two methods can be described, where the cluster amplitude time derivatives are taken as auxiliary conditions, leading to a shifted time-dependent Hamiltonian matrix. In this framework, TDCC has a shifted Hamiltonian with a block upper triangular structure, explaining the correct scaling properties of the method.
\end{abstract}

\maketitle

\section{Introduction}
With recent developments in the shaping and amplification of laser pulses, the production of short and strong pulses can now be realized in several frequency domains~\cite{RevModPhys.72.545,Seddon_2017,Duris2020}. The progress sparks further interest in the dynamical and non-linear response of molecules to strong fields, which can involve a high number of quantum states~\cite{PhysRevLett.121.173005}. This since many of the states that are inaccessible by a single-photon transition, either energetically or by symmetry selection rules, can be accessed by a multiphoton transition~\cite{doi:10.1146/annurev.physchem.040808.090427,Kulander2006}.

The ultrafast non-linear response of molecules to strong fields can give an extended degree of dynamic control of chemical reactions~\cite{Bayer_2008,doi:10.1246/bcsj.20200158,C5CP00627A}. It can also reveal information about the system that is inaccessible in weaker fields, which can be used for improving the imaging of different reaction stages~\cite{doi:10.1146/annurev.physchem.040808.090427,doi:10.1246/bcsj.20200158}. That said, the involved coupling between the numerous affected states can lead to an intricate relationship between the shape and strength of laser pulses and the molecular response, which calls for the interpretation by appropriate quantum chemistry methods~\cite{https://doi.org/10.1002/wcms.1430}.

Both the accuracy and computational complexity of quantum chemistry methods often increase with the order of approximation of the particle correlations in the system~\cite{https://doi.org/10.1002/wcms.1430,PhysRevA.93.013426} and the size of the finite basis set~\cite{doi:10.1021/acs.jctc.8b00656}, but the accuracy also depends heavily on the mathematical structure of the method. Considerable effort is spent on constructing the most well-behaved methods for a given order of computational complexity, with respect to both numerical stability and correspondence to experimental results for various systems. Real-time variants of quantum chemistry methods are convenient for modeling multiphoton transitions in systems~\cite{doi:10.1021/acs.chemrev.0c00223}, as expressions for the high-order frequency response can be difficult to both derive and to solve numerically.

The well-established single-reference coupled cluster (CC) hierarchy of methods often gives accurate and rapidly converging molecular properties~\cite{BARTLETT20051191} for states with weak multi-reference character~\cite{doi:10.1063/1.5039496}. An important reason for this accuracy is the physically reasonable scaling properties of the methods, even when the cluster operator is truncated. For instance, the energy of the ground state is size-extensive, meaning that it scales linearly with the number of non-interacting identical subsystems~\cite{doi:10.1146/annurev.pc.32.100181.002043}. In the equation-of-motion framework, the excitation energies are size-intensive, meaning that they do not scale with the number of non-interacting subsystems~\cite{doi:10.1063/1.458815}. In the linear response framework, which is based on time-dependent coupled cluster theory, ground state--excited state transition moments are also size-intensive~\cite{doi:10.1063/1.466321}. Truncated configuration interaction methods, on the other hand, do not possess these properties, and errors generally increase with the size of the simulated system~\cite{doi:10.1080/00268970500083952}.

Traditionally, the coupled cluster methods have almost exclusively been treated in the frequency domain, but the last decade has witnessed an increased exploration into their real-time behavior~\cite{doi:10.1021/acs.chemrev.0c00223}. As demonstrated by Pedersen and Kvaal~\cite{doi:10.1063/1.5085390} and further investigated by Kristiansen et al. ~\cite{doi:10.1063/1.5142276}, the exponential parametrization makes the standard time-dependent coupled cluster (TDCC) method inherently unstable whenever the reference determinant weight is depleted by a strong field. These instabilities can require the use of exceedingly small time steps in numerical solutions, and can at times also lead to breakdowns that cannot be solved by decreasing the time step size.

The orbital adaptive time-dependent coupled cluster (OATDCC) method, which requires the solution of an additional set of linear equations at each time step, was shown to have a greater stability region than TDCC. Nonetheless, the method still fails at higher field strengths, where the reference determinant weight can become greater than one~\cite{doi:10.1063/1.5142276}.

Variants of the time-dependent equation-of-motion coupled cluster (TD-EOM-CC) method have also been used for modeling laser-molecule interactions, but only a handful applications have included the full non-linear real-time propagation of the laser-driven electron dynamics~\cite{doi:10.1021/jp107384p,doi:10.1080/00268976.2012.675448,doi:10.1021/acs.chemrev.0c00223,PhysRevA.105.023103}. In these cases, the TD-EOM-CC equations were expressed in the basis obtained by diagonalizing the field-free equation-of-motion coupled cluster Hamiltonian. We instead express the equations in the elementary basis, leading to equations that are simple to implement and have computational and memory requirements that scale more favorably with respect to system size than the full diagonal basis equations. This makes the formulation particularly useful for assessing the short-time and non-linear behavior of the TD-EOM-CC method.

The paper is organized as follows. In \cref{sec:theory}, the TDCC and TD-EOM-CC methods are described in a general framework, and it is shown how the time derivative of the cluster amplitudes affects the analytical scaling properties of the two methods. \Cref{sec:computational} outlines the computational methods used to simulate atoms undergoing semiclassical Rabi oscillations in a resonant electromagnetic field. In \cref{sec:results}, results of the simulations are presented and discussed, including a demonstration of how the time-dependent energy scales with respect to system size in the two methods. The key findings are summarized in \cref{sec:conclusion}.

\section{Theory}\label{sec:theory}
\subsection{System}
The time-dependent system of the molecule and the external field is described by the Hamiltonian
\begin{equation}
    H(t) = H^{(0)} + V(t).
\end{equation}
The field-free molecular system is described by the Hamiltonian $H^{(0)}$, and the interaction between the molecular system and the external field is described by $V(t)$. We describe the interaction semi-classically, in the dipole approximation and length gauge. This gives $V(t) = - \vb*{\mu}\cdot\vb*{\mathcal{E}}(t)$, where $\vb*{\mu}$ is the electric dipole moment vector and $\vb*{\mathcal{E}}(t)$ the time-dependent electric field vector. The system is also treated within the Born-Oppenheimer approximation, with fixed nuclei.

\subsection{Time dependence in coupled cluster methods}
Coupled cluster ket and bra vectors that encompass both the TDCC and TD-EOM-CC parametrizations can be defined as
\begin{align}
    \label{eq:ket}
    \ket{\Psi(t)} &= e^{T(t)}R(t)\ket{\HF}, \\
    \label{eq:bra}
    \bra*{\widetilde{\Psi}(t)} &= \bra{\HF}L(t)e^{-T(t)},
\end{align}
where the cluster operator
\begin{equation}
    T(t) = \sum_{\kappa\geq 0} \tau_\kappa t_\kappa(t)
\end{equation}
and the right and left EOM-CC operators
\begin{equation}
    \label{eq:rightleft}
    R(t) = \sum_{\kappa\geq 0}\tau_{\kappa}r_\kappa(t) \qc L(t) = \sum_{\kappa\geq 0} l_\kappa(t)\widetilde{\tau}_{\kappa}^\dagger.
\end{equation}
The operators with index $0$ are the unit operator
\begin{equation}
    \label{eq:tau0}
    \tau_0 = \widetilde{\tau}_0^\dagger = \mathbbm{1},
\end{equation}
and the operators $\tau_\mu$ and $\widetilde{\tau}_\mu^\dagger$, where $\mu > 0$, excite and deexcite electrons between occupied and virtual Hartree-Fock molecular orbitals, respectively,
\begin{alignat}{4}
    &\tau_\mu\ket{\HF} &&= \ket{\mu} \qc &&\bra{\HF}\widetilde{\tau}_\mu^\dagger &&= \bra{\widetilde{\mu}}, \\
    &\widetilde{\tau}_\mu^\dagger\ket{\HF} &&= 0 \qc &&\bra{\HF}\tau_\mu &&= 0.
\end{alignat}
The operators are chosen so that the excited determinants are biorthonormal,
\begin{equation}
    \label{eq:subsystembiorthonormal}
    \bra{\widetilde{\kappa}}\ket{\lambda}=\delta_{\kappa\lambda} \qc \kappa \geq 0 \qc \lambda \geq 0,
\end{equation}
where $\delta_{\kappa\lambda}$ is the Kronecker delta. The field-free coupled cluster ground state can be defined by setting the amplitudes $r_\mu^{(0)}= 0$ and $r_0^{(0)} =  l_0^{(0)} = 1$, and letting the ground state cluster amplitudes $t_\mu^{(0)}$ and left vector $l_\mu^{(0)}$ be determined as solutions of the field-free ground state equations
\begin{align}
    \label{eq:omega}
    &\bra{\widetilde{\mu}}e^{-T^{(0)}}H^{(0)}e^{T^{(0)}}\ket{\HF} = 0, \\
    &\bigg(\bra{\HF} + \sum_{\mu>0}l_\mu^{(0)}\bra{\widetilde{\mu}}\bigg)\comm*{e^{-T^{(0)}}H^{(0)}e^{T^{(0)}}}{\tau_\nu}\ket{\HF} = 0.
\end{align}
For simplicity, we also set the undetermined phase-related cluster amplitude $t_0^{(0)} = 0$.

The equations for the time dependence of the parameters of \cref{eq:rightleft} can be derived from the right and left time-dependent Schr\"{o}dinger equations (TDSEs)
\begin{align}
    \label{eq:righttdse}
    i\dv{t}\ket{\Psi(t)} &= H(t)\ket{\Psi(t)}, \\
    \label{eq:lefttdse}
    -i\dv{t}\bra*{\widetilde{\Psi}(t)} &= \bra*{\widetilde{\Psi}(t)}H(t).
\end{align}
Inserting \cref{eq:ket} into \cref{eq:righttdse} before projecting onto $\bra{\widetilde{\kappa}}e^{-T(t)}$, and likewise inserting \cref{eq:bra} into \cref{eq:lefttdse} before projecting onto $e^{T(t)}\ket{\lambda}$, the following matrix-vector TDSEs are obtained
\begin{align}
    \label{eq:rightderivative}
    i\dv{r_\kappa(t)}{t} &= \sum_{\lambda\geq 0} \widetilde{H}_{\kappa\lambda}(t)r_\lambda(t), \\
    \label{eq:leftderivative}
    -i\dv{l_\lambda(t)}{t} &= \sum_{\kappa\geq 0} l_\kappa(t)\widetilde{H}_{\kappa\lambda}(t),
\end{align}
where the shifted Hamiltonian
\begin{equation}
    \label{eq:htilde}
    \widetilde{H}(t) = H(t) - i\dv{T(t)}{t}.
\end{equation}
The elements of the coupled cluster matrix $\vb*{O}(t)$ of operator $O(t)$ are given by
\begin{equation}
    \label{eq:matrixelement}
    O_{\kappa\lambda}(t) = \bra{\widetilde{\kappa}}\widebar{O}(t)\ket{\lambda},
\end{equation}
where an overbar is used to denote the similarity transformation by the exponentiated time-dependent cluster operator,
\begin{equation}
    \label{eq:similarity}
    \widebar{O}(t) = e^{-T(t)}O(t)e^{T(t)}.
\end{equation}

In TD-EOM-CC, the time derivatives of the cluster amplitudes are given by
\begin{equation}
    \label{eq:tdeomccderivative}
    i\dv{t_\kappa(t)}{t} = 0,
\end{equation}
while in TDCC, the derivatives are given by
\begin{equation}
    \label{eq:tdccderivative}
    i\dv{t_\kappa(t)}{t} = \bra{\widetilde{\kappa}}\widebar{H}(t)\ket{\HF}.
\end{equation}
The resolution of identity $\mathbbm{1}=\sum_{\eta\geq0}\ket{\eta}\bra{\widetilde{\eta}}$ and \cref{eq:htilde} can be used to rewrite the matrix elements of the shifted Hamiltonian in TDCC as
\begin{equation}
    \begin{split}
        \label{eq:tdcccommutator}
        \widetilde{H}_{\kappa\lambda}(t)
        &= \bra{\widetilde{\kappa}}\widebar{H}(t)\ket{\lambda} - \sum_{\eta\geq0}\bra{\widetilde{\kappa}}\tau_\eta\ket{\lambda}\bra{\widetilde{\eta}}\widebar{H}(t)\ket{\HF} \\
        &= \bra{\widetilde{\kappa}}\comm{\widebar{H}(t)}{\tau_\lambda}\ket{\HF}.
    \end{split}
\end{equation}

Once all time-dependent amplitudes have been found at a given point in time $t$, the time-dependent expectation values of the time-dependent operator $O(t)$ can be obtained by
\begin{equation}
    \label{eq:expval}
    \begin{split}
        \ev{O(t)} &= \bra*{\widetilde{\Psi}(t)}O(t)\ket{\Psi(t)} \\
        &= \sum_{\kappa,\lambda\geq0}l_\kappa(t)O_{\kappa\lambda}(t)r_\lambda(t) \\
        &= \vb*{l}^T(t)\vb*{O}(t)\vb*{r}(t).
    \end{split}
\end{equation}

\subsection{Scaling properties of real-time coupled cluster methods}\label{sec:scaling}
In order to theoretically investigate the scaling properties of methods based on the parametrization in \cref{eq:ket} and \cref{eq:bra}, we assume that the system is composed of non-interacting subsystems. We let $\tau_{\lambda_\Ss}$ denote an elementary excitation operator and $\widetilde{\tau}_{\kappa_{\Ss}}^\dagger$ an elementary deexcitation operator of subsystem $\Ss$. The elementary excitation and deexcitation operators of the composite system can be constructed as tensor products of all the operators of the different subsystems. Untruncated TDCC and TD-EOM-CC methods can represent all tensor products, since the excitation and deexcitation levels of the methods are not limited. In truncated methods, however, all elementary excitation and deexcitation operators that exceed a truncation level specific to the method are excluded, which can lead to errors related to the scaling from one to several subsystems.

For two subsystems $\Ss \in \{\A, \B\}$, the elementary excitation and deexcitation operators of the composite system can be constructed as the tensor products $\tau_{\lambda_\A}\otimes\tau_{\lambda_\B}$ and $\widetilde{\tau}_{\kappa_\A}^\dagger\otimes\widetilde{\tau}_{\kappa_\B}^\dagger$. We split the sets of these operators into four partitions, which we label by $0$, $\A$, $\B$ and $\AB$. The $0$ partition includes the operators that do not change the excitation level of the subsystems, $\tau_{0_\A}\otimes\tau_{0_\B}$ and $\widetilde{\tau}_{0_\A}^\dagger\otimes\widetilde{\tau}_{0_\B}^\dagger$. The $\A$ partition includes the operators that change the excitation level of subsystem $\A$ only, $\tau_{\mu_\A}\otimes\tau_{0_\B}$ and $\widetilde{\tau}_{\mu_\A}^\dagger\otimes\widetilde{\tau}_{0_\B}^\dagger$, and the $\B$ partition the operators that change the excitation level of subsystem $\B$ only, $\tau_{0_\A}\otimes\tau_{\mu_\B}$ and $\tau_{0_\A}^\dagger\otimes\tau_{\mu_\B}^\dagger$, where $\mu>0$. The $\AB$ partition includes the operators that change the excitation level of both subsystems, $\tau_{\nu_\A}\otimes\tau_{\nu_\B}$ and $\widetilde{\tau}_{\mu_\A}^\dagger\otimes\widetilde{\tau}_{\mu_\B}^\dagger$, where $\mu>0$ and $\nu>0$. Truncation can affect the $\AB$ partition, since the tensor products of the truncated subsystem operators can include excitations and deexcitations that in combination go beyond the truncation level of the method. In the following, we assess the effect this truncation has for the TDCC and TD-EOM-CC methods.

We start by assuming that the cluster amplitudes corresponding to the operators $\tau_{\nu_\A}\otimes\tau_{\nu_\B}$ are zero at a given time $t$. The cluster operator $T(t)$ can then be written as the tensor sum,
\begin{align}
    \label{eq:ttensorsum}
    T(t) = T_\A(t)\otimes I_\B + I_\A \otimes T_\B(t),
\end{align}
where $T_\Ss(t)$ is the cluster operator for subsystems $\Ss$. Since operators on non-interacting subsystems commute, we have that
\begin{equation}
    \label{eq:exponentiatedclustersum}
    e^{\pm\big(T_\A(t)\otimes I_\B + I_\A \otimes T_\B(t)\big)} = e^{\pm T_\A(t)}\otimes e^{\pm T_\B(t)}.
\end{equation}
We furthermore let $O(t)$ be any operator that operates independently on the two subsystems and thus can be written as the tensor sum
\begin{equation}
    \label{eq:otensorsum}
    O(t) = O_\A(t)\otimes I_\B + I_\A \otimes O_\B(t),
\end{equation}
where $O_\A(t)$ and $O_\B(t)$ are subsystem operators. \Cref{eq:exponentiatedclustersum} then implies that the similarity transformed operator in \cref{eq:similarity} can be written as the tensor sum
\begin{gather}
    \label{eq:operatorbarsplit}
    \begin{split}
        \widebar{O}(t)
        &= e^{-T_\A(t)}O_\A(t)e^{T_\A(t)}\otimes I_\B \\
        &\phantom{{}={}} + I_\A\otimes e^{-T_\B(t)}O_\B(t)e^{T_\B(t)} \\
        &= \widebar{O}_\A(t)\otimes I_\B + I_\A\otimes\widebar{O}_\B(t).
    \end{split},
\end{gather}
which does not contain terms where both two subsystems are excited simultaneously.

We furthermore assume that the time-dependent Hamiltonian $H(t)$ can be written on the form of \cref{eq:otensorsum}. In TDCC, the time derivative of the cluster amplitudes in \cref{eq:tdccderivative} can for the $\AB$ partition be written as
\begin{equation}
    \begin{split}
        i\dv{t_{\mu_\A\mu_\B}(t)}{t} &= \big(\bra{\widetilde{\mu}_A}\otimes\bra{\widetilde{\mu}_\B}\big) \\
        &\phantom{{}={}}\times\big(\widebar{H}_\A(t)\otimes I_\B + I_\A\otimes\widebar{H}_\B(t)\big) \\
        &\phantom{{}={}}\times\big(\ket{\HF_\A}\otimes\ket{\HF_\B}\big) \\
        &= 0.
    \end{split}
\end{equation}
As long as \cref{eq:ttensorsum} holds at the initial time, it will thus in TDCC also hold for later times. This is also trivially the case for TD-EOM-CC, since the time derivative given by \cref{eq:tdeomccderivative} is zero for all cluster amplitudes.

We let the subscript $\parallel$ denote vectors where the elements that truncated methods fail to represent have been set to zero. Accordingly, the right and left transpose vectors $\vb*{r}_\parallel$ and $\vb*{l}_\parallel^T$ are the truncated counterparts of the untruncated vectors $\vb*{r}$ and $\vb*{l}^T$, and have the following representation in the partitioned bases
\begin{equation}
    \vb*{r}_\parallel = 
    \begin{pmatrix}
        r_0 \\
        \vb*{r}_\A \\
        \vb*{r}_\B \\
        (\vb*{r}_\AB)_\parallel
    \end{pmatrix}
    \qc
    \vb*{l}_\parallel^T = 
    \begin{pmatrix}
        l_0 &
        \vb*{l}_\A &
        \vb*{l}_\B &
        (\vb*{l}_\AB)_\parallel
    \end{pmatrix},
\end{equation}
where only the $\AB$ partitions can be affected by the truncation. Furthermore, $\vb*{O}_\parallel$ is the truncated counterpart of the operator matrix $\vb*{O}$, which is the projection of \cref{eq:operatorbarsplit} onto the tensor product bases. In the partitioned bases, the matrix has the representation
\begin{equation}
    \label{eq:partitionedoperator}
    \vb*{O}_\parallel
    =
    \begin{pmatrix}
        O_{0\,0} & \vb*{O}_{0\,\A} & \vb*{O}_{0\,\B} & \vb*{0} \\
        \vb*{O}_{\A\,0} & \vb*{O}_{\A\,\A} & \vb*{0} & (\vb*{O}_{\A\,\AB})_\parallel \\
        \vb*{O}_{\B\,0} & \vb*{0} & \vb*{O}_{\B\,\B} & (\vb*{O}_{\B\,\AB})_\parallel \\
        \vb*{0} & (\vb*{O}_{\AB\,\A})_\parallel & (\vb*{O}_{\AB\,\B})_\parallel & (\vb*{O}_{\AB\,\AB})_\parallel
    \end{pmatrix}.
\end{equation}
The block matrix maps partitions of right and left transpose vectors to partitions where the numbers of excited subsystems have changed by at most one. Thus, a single matrix transformation is not enough to map between the $0$ and $\AB$ partitions, and the expectation value involving the ground state right vector $\vb*{r}_\parallel^{(0)} = (1, \vb*{0}, \vb*{0}, \vb*{0})^T$ is unaffected by product basis truncation,
\begin{equation}
    \begin{split}
        \ev{O}_\parallel &= \vb*{l}_\parallel^T\vb*{O}_\parallel\vb*{r}_\parallel^{(0)} \\
        &= l_0^TO_{0\,0} +  \vb*{l}_\A^T\vb*{O}_{\A\,0} + \vb*{l}_\B^T\vb*{O}_{\B\,0} \\
        &= \ev{O}.
    \end{split}
\end{equation}
For second- and higher-order transformations, all partitions of the transformed right and left transpose vectors can be affected by the truncation.

We now assess the effect of transformation by an operator matrix $\vb*{T}_\parallel$ with the following block upper triangular structure
\begin{equation}
    \label{eq:blocktriangular}
    \vb*{T}_\parallel
    =
    \begin{pmatrix}
        T_{0\,0} & \vb*{T}_{0\,\A} & \vb*{T}_{0\,\B} & \vb*{0} \\
        \vb*{0} & \vb*{T}_{\A\,\A} & \vb*{0} & (\vb*{T}_{\A\,\AB})_\parallel \\
        \vb*{0} & \vb*{0} & \vb*{T}_{\B\,\B} & (\vb*{T}_{\B\,\AB})_\parallel \\
        \vb*{0} & \vb*{0} & \vb*{0} & (\vb*{T}_{\AB\,\AB})_\parallel
    \end{pmatrix}.
\end{equation}
The block matrix $\vb*{T}$ maps partitions of right vectors to themselves and to partitions where the numbers of excited subsystems have decreased by one. Thus, the $0$ partition of the ground state right vector $\vb*{r}^{(0)}$ is only mapped to itself under successive transformations by block upper triangular matrices,
\begin{equation}
    \label{eq:triangularright}
    \begin{split}
        \vb*{T}_\parallel^{\prime\cdots\prime}\cdots\vb*{T}_\parallel'\vb*{T}_\parallel\vb*{r}_\parallel^{(0)} &=
        \begin{pmatrix}
            (\vb*{T}^{\prime\cdots\prime}\cdots\vb*{T}'\vb*{T}\vb*{r}^{(0)})_0 \\
            \vb*{0} \\
            \vb*{0} \\
            \vb*{0}
        \end{pmatrix},
    \end{split}
\end{equation}
where all truncated matrices $\vb*{T}_\parallel$ have been replaced by $\vb*{T}$ whenever the truncated $\AB$ partition does not make any contribution. We can see that the truncation does not affect the transformed vector in \cref{eq:triangularright} at all, while it affects all partitions of right vectors transformed by two or more matrices with the structure of $\vb*{O}$. Furthermore, $\vb*{T}$ maps partitions of left transpose vectors to themselves and to partitions where the numbers of excited subsystems have increased by one. Thus, the $\AB$ partition of left transpose vectors is only mapped to itself under successive transformations by block upper triangular matrices, and
\begin{equation}
    \label{eq:leftvector}
    \vb*{l}_\parallel^T\vb*{T}_\parallel\vb*{T}_\parallel'\cdots\vb*{T}_\parallel^{\prime\cdots\prime} = \begin{pmatrix}
        (\vb*{l}^T\vb*{T}\vb*{T}'\cdots\vb*{T}^{\prime\cdots\prime})_0 \\
        (\vb*{l}^T\vb*{T}\vb*{T}'\cdots\vb*{T}^{\prime\cdots\prime})_\A \\
        (\vb*{l}^T\vb*{T}\vb*{T}'\cdots\vb*{T}^{\prime\cdots\prime})_\B \\
        (\vb*{l}_\parallel^T\vb*{T}_\parallel\vb*{T}_\parallel'\cdots\vb*{T}_\parallel^{\prime\cdots\prime})_\AB
    \end{pmatrix}^T,
\end{equation}
We can see that the truncation only affects the $\AB$ partitions of the transformed left transpose vector in \cref{eq:leftvector}, while it affects all partitions of left transpose vectors transformed by two or more matrices with the structure of $\vb*{O}$.

In the truncated product basis, the exact solutions of the right and left matrix TDSEs in \cref{eq:rightderivative} and \cref{eq:leftderivative} can be given by
\begin{equation}
    \label{eq:}
    \vb*{r}_\parallel(t) = \vb*{U}_\parallel(t, t_0)\vb*{r}_\parallel(t_0) \qc \vb*{l}_\parallel^T(t) = \vb*{l}_\parallel^T(t_0)\vb*{U}_\parallel(t_0, t),
\end{equation}
where
\begin{align}
    \begin{split}
        \vb*{U}_\parallel(t, t_0) &= \vb*{1}_\parallel -i \int_{t_0}^t\dd{t'}\widetilde{\vb*{H}}_\parallel(t') \\
        &\phantom{{}={}}+(-i)^2\int_{t_0}^t\dd{t'}\int_{t_0}^{t'}\dd{t''}\widetilde{\vb*{H}}_\parallel(t')\widetilde{\vb*{H}}_\parallel(t'')+\cdots.
    \end{split}
\end{align}
Under the conditions that the matrix $\vb*{\widetilde{H}}(t)$ has the block upper triangular structure of $\vb*{T}$ in \cref{eq:blocktriangular}, and that the right vector starts out as the ground state vector $\vb*{r}(t_0) = \vb*{r}^{(0)}$, \cref{eq:triangularright} implies that
\begin{equation}
    \vb*{r}(t) = \vb*{r}_\parallel(t) =
    \begin{pmatrix}
        \big(\vb*{U}(t,t_0)\vb*{r}(t_0)\big)_0 \\
        \vb*{0} \\
        \vb*{0} \\
        \vb*{0}
    \end{pmatrix},
\end{equation}
where $\vb*{U}(t,t_0)$ denotes the time evolution operator with the untruncated time-dependent Hamiltonian matrix $\widetilde{\vb*{H}}(t)$. Furthermore, \cref{eq:leftvector} implies that
\begin{equation}
    \vb*{l}_\parallel(t) =
    \begin{pmatrix}
        \big(\vb*{l}^T(t_0)\vb*{U}(t_0, t)\big)_0 \\
        \big(\vb*{l}^T(t_0)\vb*{U}(t_0, t)\big)_\A \\
        \big(\vb*{l}^T(t_0)\vb*{U}(t_0, t)\big)_\B \\
        \big(\vb*{l}_\parallel^T(t_0)\vb*{U}_\parallel(t_0, t)\big)_\AB
    \end{pmatrix}.
\end{equation}
Under these conditions, time-dependent expectation values are not affected by the truncation of the product basis
\begin{equation}
    \label{eq:triangularexpectation}
    \begin{split}
        \ev{O(t)}_\parallel &= \vb*{l}_\parallel^T(t)\vb*{O}_\parallel(t)\vb*{r}_\parallel(t) \\
        &= \big(\vb*{l}^T(t_0)\vb*{U}(t_0, t)\big)_0O_{0\,0}(t)\big(\vb*{U}(t,t_0)\vb*{r}(t_0)\big)_0 \\
        &\phantom{{}={}}+ \big(\vb*{l}^T(t_0)\vb*{U}(t_0, t)\big)_\A\vb*{O}_{\A\,0}(t)\big(\vb*{U}(t,t_0)\vb*{r}(t_0)\big)_0 \\
        &\phantom{{}={}}+ \big(\vb*{l}^T(t_0)\vb*{U}(t_0, t)\big)_\B\vb*{O}_{\B\,0}(t)\big(\vb*{U}(t,t_0)\vb*{r}(t_0)\big)_0 \\
        &=\ev{O(t)},
    \end{split}
\end{equation}
and thus behave correctly when the system scales from one to two non-interacting subsystems. Furthermore, the expectation values can be shown to scale correctly to any number of non-interacting subsystems by repeatedly splitting one of the remaining composite subsystems in two before repeating the above arguments.

From \cref{eq:tdcccommutator}, we can see that the blocks $\widetilde{\vb*{H}}_{\A\,0}(t)$ $\widetilde{\vb*{H}}_{\B\,0}(t)$, $\widetilde{\vb*{H}}_{\AB\,\A}(t)$ and $\widetilde{\vb*{H}}_{\AB\,\B}(t)$ below the diagonal of the shifted Hamiltonian matrix are equal to zero in truncated TDCC methods. This implies that the shifted time-dependent Hamiltonian $\widetilde{\vb*{H}}(t)$ has the same triangular block structure as $\vb*{T}$ in \cref{eq:blocktriangular}, and the method has the correct scaling properties when the system starts out in the ground state, in accordance with \cref{eq:triangularexpectation}. Note that the amplitude derivative $i\dv{t_0(t)}{t} = \bra{\HF}\widebar{H}(t)\ket{\HF}$ in \cref{eq:tdccderivative} also implies that $\widetilde{H}_{00}(t)$ is zero, but this condition is not needed for the correctness of the scaling properties of truncated TDCC methods.

In general, however, time-dependent expectation values do not have to be same in truncated and untruncated methods,
\begin{equation}
    \label{eq:generalexpectation}
    \ev{O(t)}_\parallel = \vb*{l}_\parallel^T(t)\vb*{O}_\parallel(t)\vb*{r}_\parallel(t) \neq \vb*{l}^T(t)\vb*{O}(t)\vb*{r}(t) =\ev{O(t)}.
\end{equation}
As stated in \cref{eq:tdeomccderivative}, the time derivatives of the cluster amplitudes are zero in TD-EOM-CC, and therefore $\widetilde{\vb*{H}}(t) = \vb*{H}(t)$. The interaction term $\vb*{V}(t)$ of the Hamiltonian is in general not block upper triangular, and can map partitions of right and left transpose vectors in the same way as $\vb*{O}$. The truncation of the product basis can thus affect all partitions of the right and left vectors, and time-dependent expectation values of TD-EOM-CC are in general misrepresented when the product basis is truncated, in accordance with \cref{eq:generalexpectation}. However, the field-free term $\vb*{H}_0$ of the time-dependent Hamiltonian has the block upper triangular structure of $\vb*{T}$, and the truncation of the product basis does thus not affect expectation values when there is no interaction with the external field.

\section{Computational details}\label{sec:computational}
In order to numerically assess the behavior of TD-EOM-CC, we implement the method described by the differential equations \cref{eq:rightderivative}, \cref{eq:leftderivative} and \cref{eq:tdeomccderivative} in the spin-adapted elementary basis, and the expectation value expression \cref{eq:expval} in a development version of the $e^T$ program~\cite{Folkestad2020}. We furthermore use the existing spin-adapted ground state and TDCCSD methods in $e^T$ 1.0~\cite{Folkestad2020,PhysRevA.102.023115}. The methods are used to calculate the interaction of atoms with the electromagnetic field represented by the electric field
\begin{equation}
    \vb*{\mathcal{E}}(t) = \mathcal{E}_0\vb*{\epsilon}\cos(\omega_0(t-t_0) + \phi)f(t)
\end{equation}
where $\epsilon_0$ is the peak field strength, $\vb*{\epsilon}$ the polarization, $\omega_0$ the carrier frequency, $\phi$ the carrier-envelope phase and $f(t)$ the envelope of the field. The envelope is given the functional form
\begin{equation}
    f(t) =
    \begin{cases}
        0 \qc & t < a, \\
        \sin[2](\frac{2\pi(t-a)}{4(b-a)}) \qc & a \le t \le b, \\
        1 \qc & t > b,
    \end{cases}
\end{equation}
which increases from zero to one in the interval from $a$ to $b$.

The cc-pVTZ basis set is used for the helium and beryllium atoms in the simulations. The laser field is given a carrier frequency of \SI{1.88043392}{\au}, corresponding to the transition between the ground $0\,{}^1\mathrm{S}_0$ state and the first dipole-allowed excited $2\,{}^1\mathrm{P}_1$ state of helium. The field is furthermore given a peak field strength of \SI{2.5e-2}{\au}, a polarization in the $z$-direction and a carrier-envelope phase of $\phi=- \pi/2$. The envelope of the field is set to increase from $a = \SI{0}{\au}$ of time until $b=\num{25}$ optical cycles (\SI{\approx83.534}{\au}). The envelope gives the laser field a narrow bandwidth, centered around the $0\,{}^1\mathrm{S}_0$--$2\,{}^1\mathrm{P}_1$ resonance, which ensures that the state is kept in a time-dependent superposition dominated by the two states. The integration of the time-dependent differential equations is done using the Dormand-Prince method of order 5(4) with the adaptive time stepping procedure described in Appendix B of Ref.~\cite{PhysRevA.105.023103}. The initial time step size is set to \SI{1e-2}{\au}, and the maximum and minimum values of the estimated error are set to \SI{1e-7}{\au} and \SI{1e-9}{\au}, respectively.

\section{Results and discussion}\label{sec:results}

\begin{figure}
    \centering
    \includegraphics[width=3.375in]{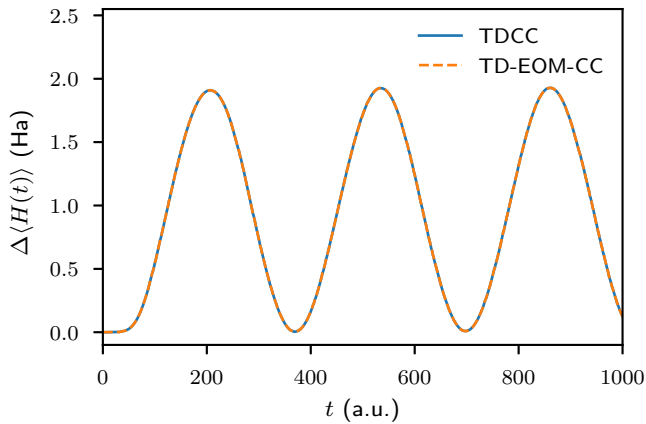}
    \caption{Time-dependent TDCC and TD-EOM-CC simulations of a single helium atom in a slowly ramped laser field. The time-dependent energy difference $\Delta\ev{H(t)}=\ev{H(t)}-E_0$ of each simulation is shown as a function of time.}
    \label{fig:heliumtdcctdeomcc}
\end{figure}

\subsection{Simulating single-subsystem Rabi oscillations with TDCC and TD-EOM-CC}
For a single helium atom, the TDCCSD and TD-EOM-CCSD methods can describe all possible excitations of the reference determinant, and the time-dependent observables are thus analytically equal for the two methods. In \cref{fig:heliumtdcctdeomcc}, we demonstrate that this is also the case numerically for the time-dependent excitation energy, as the results are equal for the two methods. The excitation energy starts out at zero, and periodically increases and decreases as a function of time, illustrating that the system undergoes Rabi oscillation between the $0\,{}^1\mathrm{S}_0$ and $2\,{}^1\mathrm{P}_1$ states. TDCC is known to be numerically unstable when the weight of the reference determinant approaches zero~\cite{doi:10.1063/1.5085390,doi:10.1063/1.5142276}, but we observe that the method can be used for simulating an essentially complete population inversion for the single helium atom.

\begin{figure*}
    \centering
    \includegraphics[width=6.75in]{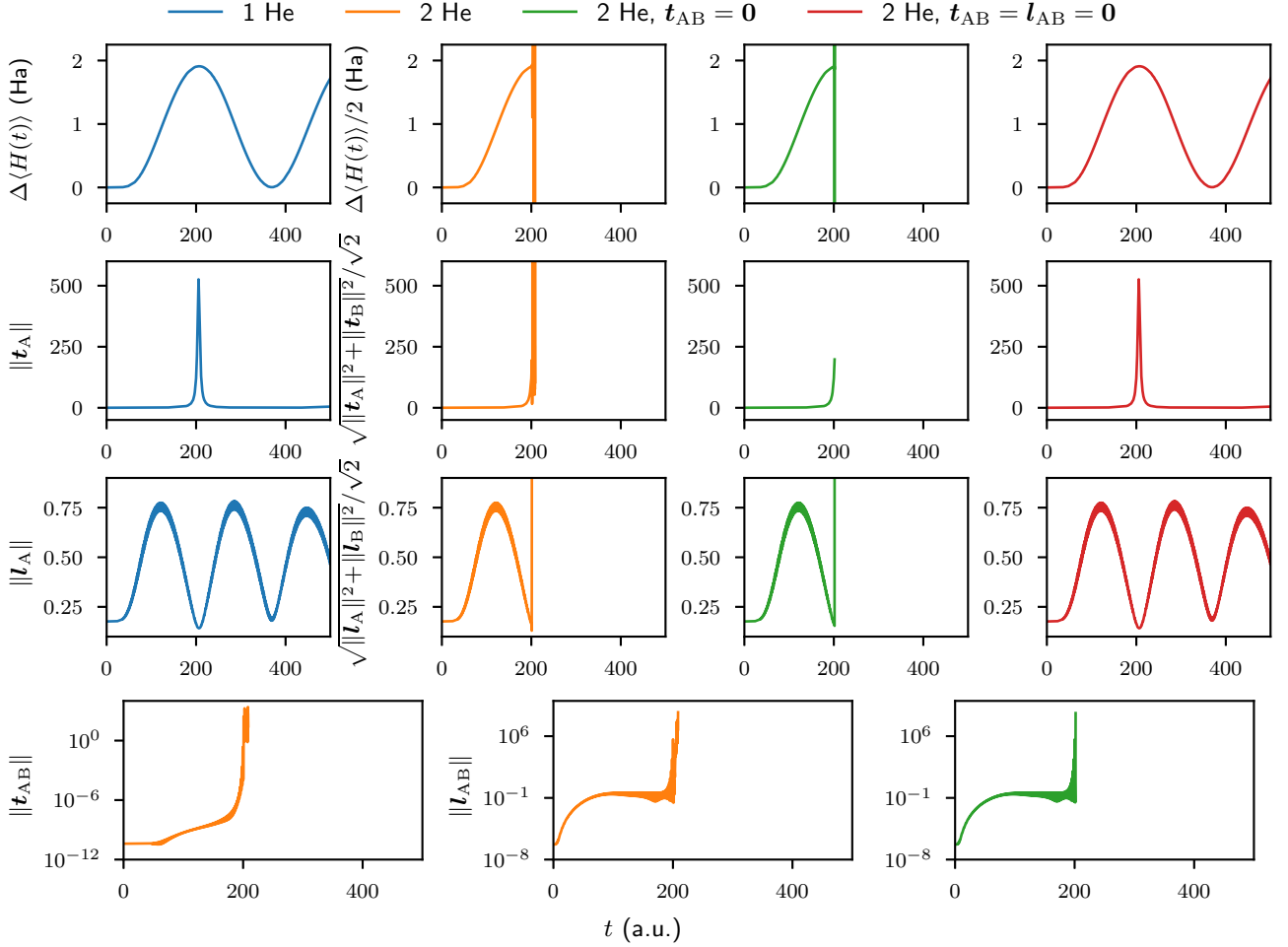}
    \caption{TDCC simulations of one helium atom (1 He) and two non-interacting helium atoms (2 He) in a slowly ramped laser field. The two helium atoms are simulated with regular TDCC, but also with TDCC where the initial values and time derivatives of the cluster amplitudes in the $\AB$ partition are set to zero ($\vb*{t}_{\AB}=0$), and TDCC where the initial values and time derivatives of both the cluster and left vector amplitudes in the $\AB$ partition are set to zero ($\vb*{t}_{\AB}=0$, $\vb*{l}_{\AB}=0$). In the top row of panels, the time-dependent energy differences $\Delta\ev{H(t)}=\ev{H(t)}-E_0$ are shown, where the two-helium results are scaled by $1/2$ and $E_0$ is the ground state energy. In the next two rows, the norms of the amplitude and left vector partitions corresponding to an excitation of a single subsystem are shown, where the two-helium results are scaled by $1/\sqrt{2}$. In the bottom panel row, the norms of the amplitude and left vectors corresponding to an excitation of two subsystems are shown.}
    \label{fig:tdccenergynorms}
\end{figure*}

\subsection{Simulating collective Rabi oscillations with TDCC}
To numerically investigate the scaling properties of TDCC, we compare the results from the single-helium simulation with results from simulations of two effectively non-interacting helium atoms, where one is placed at the origin and the other at \SI{1000}{\angstrom} on the $x$-axis, respectively. The time-dependent excitation energy of each simulation is shown in the top row of \cref{fig:tdccenergynorms}. Until around \SI{200}{\au} of time, we can see that the excitation energy of the two-helium TDCC simulation, shown in the second column, is two times the excitation energy of the single-helium simulation in the first column. This is in accordance with the theory in \cref{sec:theory}, and implies that TDCCSD treats the correlation exactly in this interval, even though the system has four electrons. We also note that the combined single-subsystem norms $\sqrt{\norm{\vb*{t}_\A}^2+\norm{\vb*{t}_\B}^2}$ and $\sqrt{\norm{\vb*{l}_\A}^2+\norm{\vb*{l}_\B}^2}$ of the two-helium calculations, shown in the second and third rows, respectively, are $\sqrt{2}$ times the respective single-helium norms $\norm{\vb*{t}_\A}$ and $\norm{\vb*{l}_\A}$ in the same interval.

After around \SI{200}{\au} of time, the two-helium solution in the second column of \cref{fig:tdccenergynorms} breaks down, and the values of the two-helium excitation energies and norms blow up. Note that the norm of the $\AB$ partition of the cluster amplitudes of the regular TDCC calculation, shown in the bottom left panel, starts out as very small (less than \num{1e-10}), but grows continually during the propagation. As the solution approaches the breakdown at around \SI{200}{\au} of time, the norm of the $\AB$ partition of the cluster amplitudes grows very rapidly. Likewise, the left vector norm of the regular TDCC calculation is also small at the start of the propagation, and blows up at around \SI{200}{\au} of time. We have performed calculations with various separations or the two helium atoms up to \SI{1e6}{\angstrom}, and the solutions still display the same behavior.

Analytically, the TDCCSD observables should have the correct scaling behavior, as demonstrated in \cref{sec:scaling}, but this is clearly not the case in our simulations. We argue that this discrepancy is due to two effects: the growth of the $\AB$ partition of the time-dependent amplitudes due to the sensitivity to small deviations from zero in the $\AB$ partition of the initial cluster amplitudes and time-dependent Hamiltonians in numerical simulations, and the blowup of the $\AB$ partition of the left vector.

In the third and fourth column of \cref{fig:tdccenergynorms}, results from modified TDCC simulations of two helium atoms are shown, where one atom is placed at the origin and the other at \SI{1000}{\angstrom} on the $x$-axis. For the results in the third column, the initial values and derivatives of the cluster amplitudes in the $\AB$ partition are set to zero. The excitation energy and norms of the remaining amplitudes still blow up around \SI{200}{\au} of time. Note that the cluster amplitudes are independent of the left vector amplitudes, and the cluster amplitudes in the $\AB$ partition would blow up regardless of the value of the left vector elements. For the results in the fourth column, the initial values and derivatives of the left vector amplitudes in the $\AB$ partition, which do not enter in TDCC expectation value expressions, are also set to zero. In this case, the numerical integration completes, and the two-helium results are equal to the single-helium results apart from the scaling factors of $2$ and $\sqrt{2}$ for the energy and single-subsystem norms, respectively.

\begin{figure}
    \centering
    \includegraphics[width=3.375in]{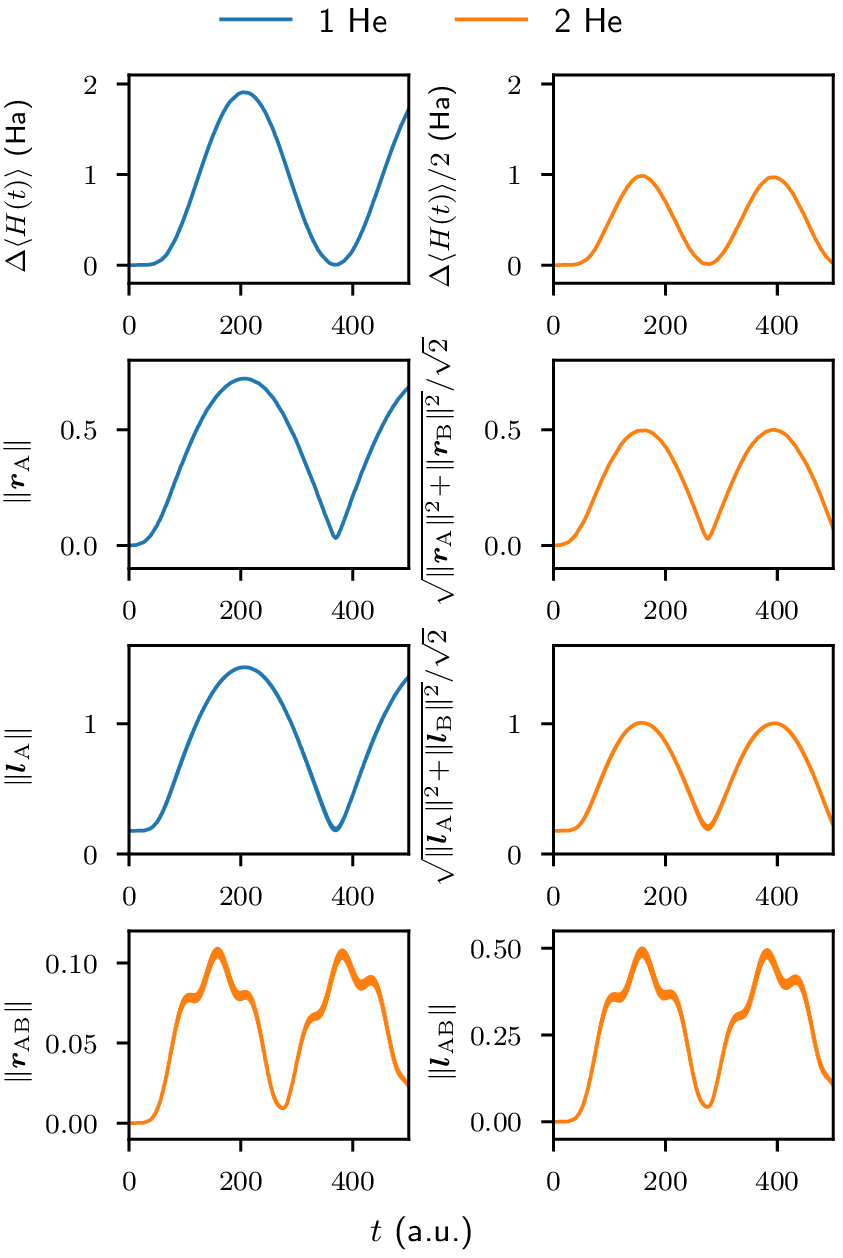}
    \caption{TD-EOM-CC simulations of one helium atom (1 He) and two non-interacting helium atoms (2 He) in a slowly ramped external field. In the first row of panels, the time-dependent energy differences $\Delta\ev{H(t)}=\ev{H(t)}-E_0$ are shown, where the two-helium results are scaled by $1/2$ and $E_0$ is the ground state energy. In the next two rows, the norms of the right and left vector partitions corresponding to an excitation of a single subsystem are shown, where the two-helium results are scaled by $1/\sqrt{2}$. In the fourth panel row, the norms of the right and left vectors corresponding to an excitation of two subsystems are shown.}
    \label{fig:tdeomccenergynorms}
\end{figure}

\begin{figure}
    \centering
    \includegraphics[width=3.375in]{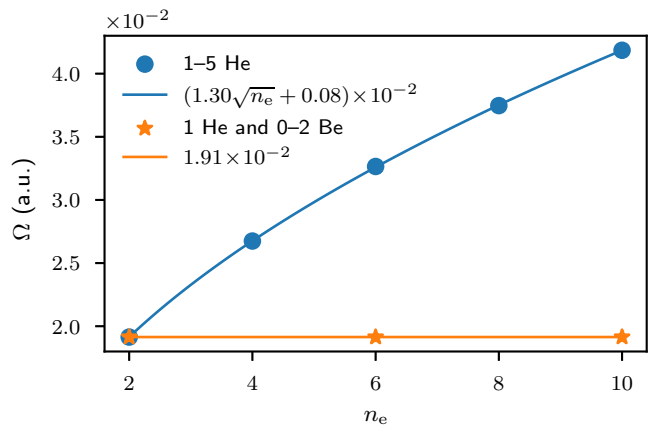}
    \caption{TD-EOM-CC calculations of helium and beryllium atoms in a slowly ramped laser field. The TD-EOM-CC Rabi frequencies $\Omega$, obtained by least-squares fitting the function $A\sin(\Omega t + \varphi) + C$ to the time-dependent energy, are given for different numbers of electrons in the system $n_\mathrm{e}$. The function $A\sqrt{n_\mathrm{e}} + C$ is least-squares fitted to the time-dependent energy for 1--5 effectively non-interacting helium atoms, giving $A = \num{1.30e-2}$ and $C = \num{8e-4}$. The constant $C$ is fitted to the Rabi frequencies for 1 helium atom and 0--2 beryllium atoms, all non-interacting, giving $C = \num{1.91e-2}$.}
    \label{fig:he_size}
\end{figure}

\subsection{Simulating collective Rabi oscillations with TD-EOM-CC}
To numerically investigate the scaling properties of TD-EOM-CC,  the interaction with the field is first calculated for two helium atoms placed at the origin and at $\SI{1000}{\angstrom}$ on the $x$-axis. The excitation energy and norms are shown together with the single-helium results in \cref{fig:tdeomccenergynorms}, where the same normalization factors has been used as for the TDCC results. The frequency of the oscillations in the scaled excitation energy and single-subsystem norms increases, and their magnitude decreases, as the number of helium atoms increases from one to two. The simulations are however numerically stable without the need for modifying the equations describing the time-dependence of the state, in contrast to the TDCC simulations.

To further investigate the scaling properties, the interaction with the field is also calculated for three to five helium atoms, and for one helium atom together with one to two beryllium atoms, where all atoms are placed \SI{1000}{\au} apart on the $x$-axis. The sinusoidal function $A\sin(\Omega t + \varphi) + C$ is least-squares fitted to the time-dependent energy between $t = 25$ optical cycles and $t = \SI{500}{\au}$. The estimated Rabi frequencies $\Omega$ of the oscillating energy are shown for all calculations in \cref{fig:he_size}. The frequencies increase with the number of non-interacting subsystem. For the purposes of quantifying the scaling properties of the Rabi frequencies, the figure also includes the function $A\sqrt{n_e}+C$ least-squared fitted to the frequencies for one to five helium atoms, where $n_e$ is the number of electrons. The goodness of the fit demonstrates that the frequency increases as the square root of the total number of helium atom electrons. As the number of non-interacting subsystems in resonance with the field increases, we can expect the frequency to falsely appear to approach infinity, meaning that the method gives a qualitatively incorrect representation of interactions occurring at multiple sites simultaneously. The figure also includes a constant $C$ fitted to the Rabi frequencies calculated with the helium atom and one to two beryllium atoms. The goodness of the fit illustrates that the frequency does not scale with the number of systems that are not in resonance with the field. The representation of the helium Rabi oscillation is therefore unaffected by the beryllium atoms, suggesting that TD-EOM-CC can represent interactions occurring at a single site of an extended quantum system.

\section{Conclusion}\label{sec:conclusion}
In this work, we have demonstrated that the TDCC and TD-EOM-CC parametrizations can be expressed in a unified theoretical framework, where the time derivatives of the cluster amplitudes are taken as auxiliary conditions. We have furthermore implemented the TD-EOM-CC method in the elementary basis, and compared the scaling properties of the TDCC and TD-EOM-CC methods through simulations of collective Rabi oscillations. We noted that the truncated TD-EOM-CC method fails to give a qualitatively correct representation of collective Rabi oscillations, as the Rabi frequency increases with the number of subsystems that are in resonance with the external field. However, we did not encounter any numerical instabilities in the TD-EOM-CC simulations, and the addition of subsystems that are not in resonance with the field did not affect the time-dependent energy. This indicates that TD-EOM-CC is suitable for simulating Rabi oscillations that occur at a single site. Although the introduction of additional Rabi oscillating subsystems negatively impacts the numerical stability of TDCC simulations, leading to solution blowup, the initial stages of the simulations display the correct scaling properties, as predicted by the theory in \cref{sec:scaling}. This supports the use of TDCC for simulating collective resonant excitations in extended systems, as long as the reference determinant weight is not completely depleted. In conclusion, we propose that further research should be dedicated to the development of approximate methods that can give a qualitatively correct description of collective Rabi oscillations without compromising numerical stability.

\begin{acknowledgments}
This research has been financially supported by the Research Council of Norway through FRINATEK project nos. 263110 and 275506, and computing resources have been provided by Sigma2 AS through project no. NN2962K. The authors would like to thank Alice Balbi for useful discussions. 
\end{acknowledgments}

\bibliographystyle{apsrev4-2}
\bibliography{bib}

\end{document}